\documentclass[10pt,aps,prd,superscriptaddress]{revtex4}
\usepackage{epstopdf}
\usepackage{graphicx}
\def\be{\begin{equation}}
\def\ee{\end{equation}}
\def\bea{\begin{eqnarray}}
\def\eea{\end{eqnarray}}
\def\no{\nonumber}
\def\dag{\dagger}
\def\wt{\widetilde}
\usepackage{amsmath}
\usepackage[T1]{fontenc}
\usepackage{fourier}

\begin{document}

\title{Entanglement in a model for Hawking radiation: An Application of Quadratic Algebras}
\author{Bindu A. Bambah}
\email{bindusp@uohyd.ernet.in}
\affiliation{School of Physics, University of Hyderabad, Gachibowli, Hyderabad, Andhra Pradesh, India 500 046}
\author{C. Mukku}
\email{mukku@iiit.ac.in}
\affiliation{International Institute of Information Technology, Gachibowli, Hyderabad, Andhra Pradesh, India 500 032}
\author{T. Shreecharan}
\email{shreecharan@gmail.com}
\affiliation{School of Physics, University of Hyderabad, Gachibowli, Hyderabad, Andhra Pradesh, India 500 046}
\author{K. Siva Prasad}
\email{kasettisivaprasad@gmail.com}
\affiliation{School of Physics, University of Hyderabad, Gachibowli, Hyderabad, Andhra Pradesh, India 500 046}
\begin{abstract}
\begin{center}
{\small ABSTRACT}
\end{center}
Quadratic polynomially deformed $su(1,1)$ and $su(2)$ algebras are utilised in model Hamiltonians to show how the gravitational system consisting of a black hole, infalling radiation and outgoing (Hawking) radiation can be solved exactly. The models allow us to study the long-time behaviour of the black hole and  its outgoing modes. In particular, we calculate the bipartite entanglement entropies of subsystems consisting of a) infalling plus outgoing modes and b) black hole modes plus the infalling modes,using  the Janus-faced nature of the model.The long-time behaviour also gives us glimpses of modifications in the character of Hawking radiation. Lastly,  we study the phenomenon of superradiance in our model in analogy with atomic Dicke superradiance.
\end{abstract}

\maketitle

\section{Introduction}
\large
Classically, black-holes, being the final state in the evolution of massive stars,  cloak their singularities from the rest of the spacetime through event horizons and have surprisingly simple properties. Observers who live in the region exterior to the event horizon cannot observe the singularity inside the event horizon while those that have travelled inside cannot communicate with observers outside the event horizon.
Black hole solutions, which are ``eternal'', can be described at most by three parameters, their mass, angular momentum and electric charge. A rotating black hole can lose energy through an energy extraction process (Penrose process-- which from a thermodynamic perspective has been called incoming wave amplification, or superradiance with associated ``grey body'' factors for the black hole\cite{schiffer}), classically, black holes cannot decrease but can only increase the area of their event horizons (through accretion of matter).

A dramatic shift in the black-hole paradigm occured when Hawking, using earlier results that a classical black hole cannot decrease the area of its event horizon, made a remarkable prediction that from a quantum mechanical prespective, a black hole can emit radiation which an observer at infinity sees as being thermal.
 It was suggested that the process itself is through quantum fluctuations in the intense gravitational fields near the event horizon of a black hole. Quantum fluctuations create pairs of particles, which are split apart, one falling into the black hole and the other radiating away in such a manner that the black hole appears to be emitting thermal radiation at  late times\cite{Hawking}.  The energy required for this radiation is believed to come from the mass of the black hole, which eventually disappears.
Hence, from a quantum mechanical viewpoint, the area of the event horizon decreases (with a corresponding decrease in its mass) until the black hole eventually ceases to exist when its mass goes to zero.
These studies of quantum fields in interaction with black holes has led to extraordinary links between black holes, thermodynamics and information theory \cite{preskill1992}. Because of this interaction with quantum fields, a black hole can be associated with temperature related to the surface gravity of the horizon. In its simplest form for uncharged, non-rotating black holes, $T \propto 1/ M$,where M is the mass of the black hole. This inverse relationship between the mass of the black hole and its temperature gives us an important clue to its evolution. In particular, Hawking has shown that the black hole behaves as a black body at this temperature and correspondingly, an observer at infinity sees thermal radiation from the black hole. Energy conservation would then imply that as the black hole radiates energy, its mass decreases with a consequent increase in temperature. It is clearly important to understand this later stage evolution of the black hole as it eventually disappears. More recently it has been suggested that if a quantum theory of gravity is to explain Hawking radiation,  then there would be a clear violation of quantum mechanics \cite{'thooft 1985}. This is caused by the evolution of pure states (states falling into black hole) into mixed states (thermal radiation from a black hole). Notwithstanding a string field theory explanation \cite{susskind,Das}, it is important to analyse the complete evolution of a black hole through to its disappearance, while examining the states radiated out to infinity. This is only possible through simple models, where, back reaction effects can be included in later stages of the evolution where they would dominate. Of course, we would like to answer the question of the information paradox being resolved through an audit of the complete radiation released out to infinity by the black hole through its lifetime.
Since it is very difficult to actually test such cosmological predictions, intra-disciplinary links in physical systems are especially important and allow analogue laboratory models to serve as  important tools in their understanding.

Laboratory  analogues for simulating Hawking radiation are now blossoming, with the advent of new techniques in condensed matter physics and quantum optics\cite{visser}.
Such models, of course, are at the most  toy models, and examples are acoustic models in moving fluids ( in which supersonic fluid flow  generate  an acoustic analogue of a ``black hole" \cite{unruh}, and the  presence of phononic Hawking radiation from the acoustic horizon is derived), optical models which use slow light passing through a Bose Einstein condensate that is itself spinning faster than the local speed of light within to create a vortex capable of trapping the light behind an event horizon  to produce an optical black hole\cite{barcello}. Recently using pico second  pulsed lasers passing through glass, the conditions for production of a disturbance zone in an optical black hole have been produced in the laboratory. This disturbance zone exhibits the emission of analogue hawking radiation for an``optical black hole"\cite{narimanov}.
There are many other ``analogue models'' that may  shed new light on perplexing theoretical questions concerning the information paradox problem. The information flow is in principle bi-directional and sometimes insights developed within the context of general relativity can be used to understand aspects of the analogue model. We seek to analyse a  model that  keeps  the  salient features with respect to just the information loss aspect,  enabling us to see whether the thermal nature of the radiation is modified when the black hole has evolved to a Planckian size and is therefore amenable to quantization. We consider a first quantized model as an illustrative example in order to examine if there is any  non-thermal contribution to the Hawking type radiation arising from the quantization of the  analogue "black hole".

We consider two such models. The first is the model introduced by
Nation and Blencowe\cite{nb}.   The second is the Tavis-Cummings version of the Dicke model\cite{tc}.  The two mode parametric amplifier with a bilinear Hamiltonian  has traditionally been used as a toy model for hawking radiation, since it produces two mode squeezed states, which can be mapped on to the incoming and outgoing states across the black hole horizon and thus simulate Hawking radiation.   However, if  we want to consider the effect of  back reaction, we have to include  the coupling of the quantized "black hole" mode. This is achieved by using a trilinear Hamiltonian \cite{brif}. Traditionally, in quantum optics such a Hamiltonian is used to describe a quantum parametric amplifier and frequency converter with a quantized pump field. The outgoing modes are identified with the signal mode, the incoming modes are identified with the idler modes and the gravitational field of the black hole  is represented by the pump \cite{sbpv}. It can be shown that if the pump mode is treated classically, the signal mode (corresponds to the outgoing particle modes)  and the idler mode(  corresponding to ingoing particle modes) form a two-mode squeezed state, and if we trace over the idler modes, the outgoing radiation can be mapped onto Hawking radiation.   In the trilinear version, the pump mode is also quantized .  This, thus, corresponds to the quantization of the black hole.  A nice pictorial analogue can be found in \cite{nb,nation}. For our purposes in this paper, our notations are such that mode $b$ is used to represent outgoing particles (Hawking radiation), $c$ is used to represent the infalling particles and $a$ is used to represent the quantized black hole modes. Essentially, the modes $b$ and $c$ represent the pair creation process at the event horizon of the black hole.

 Nation and Blencowe have studied the entanglement properties of this system using  a short time approximation\cite{nb}. Our interest is an analysis of the complete history of the radiation and the short time approximation is insufficient for our purpose.  We are able to do a comprehensive study  by exhibiting a Quadratic Polynomial Algebra (a polynomial deformation of a Lie algebra) of the trilinear Hamiltonian in two different ways. One of which is similar to that of \cite{nb}, while the other  maps the system  onto an interacting atom-radiation system  called the Tavis-Cummings model. This Janus-like nature of the mapping, enables us to study many different aspects of Hawking radiation in analogue black-hole systems.   

\section{Classical Black Hole}
\begin{center}
\begin{figure}
\includegraphics[height=3in,width=5in]{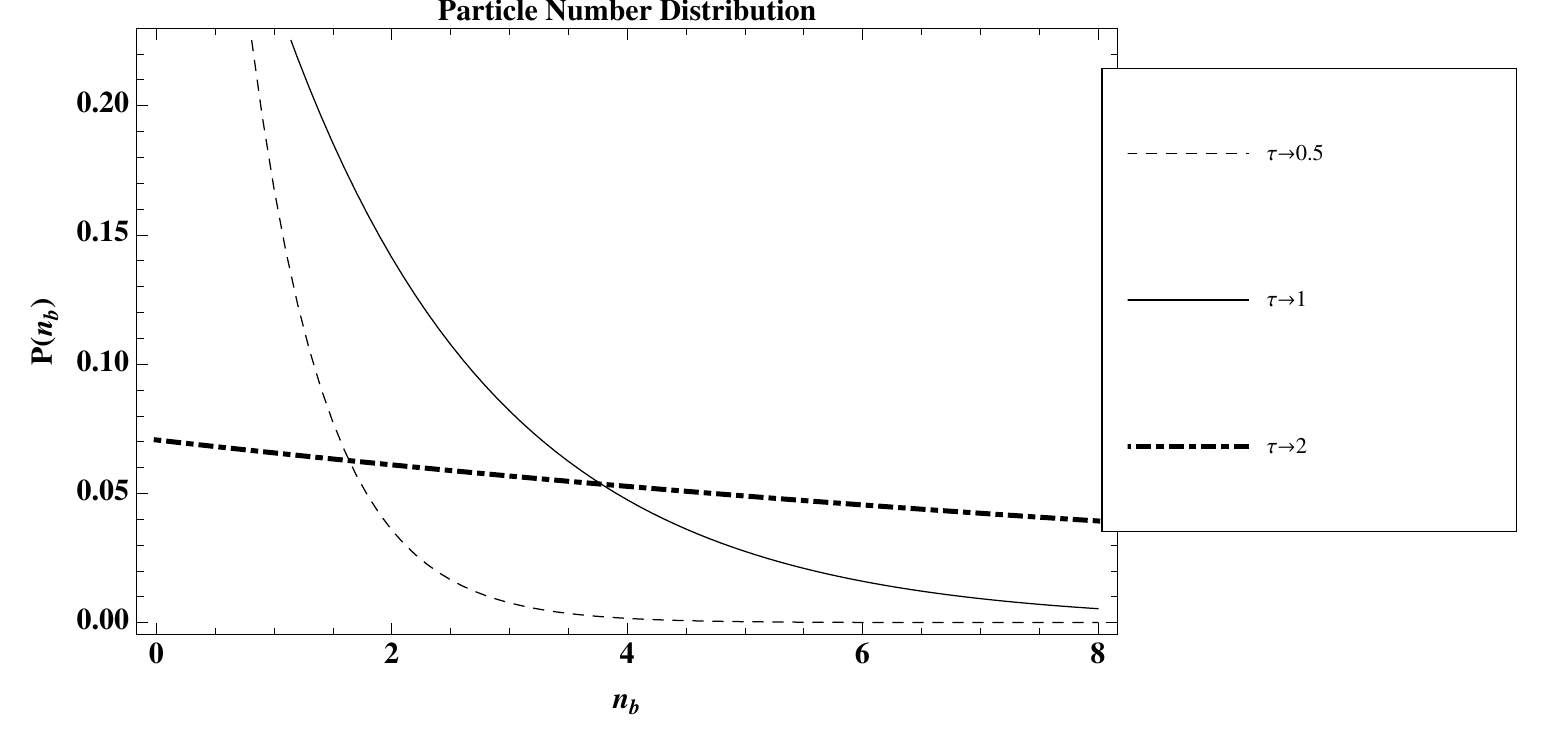}
\caption{ $P_{n_b}$ vs. $n_b$ for the outgoing modes}\label{photondist_su11}
\end{figure}
\end{center}

\begin{center}
\begin{figure}
\includegraphics[height=3in,width=4.8in]{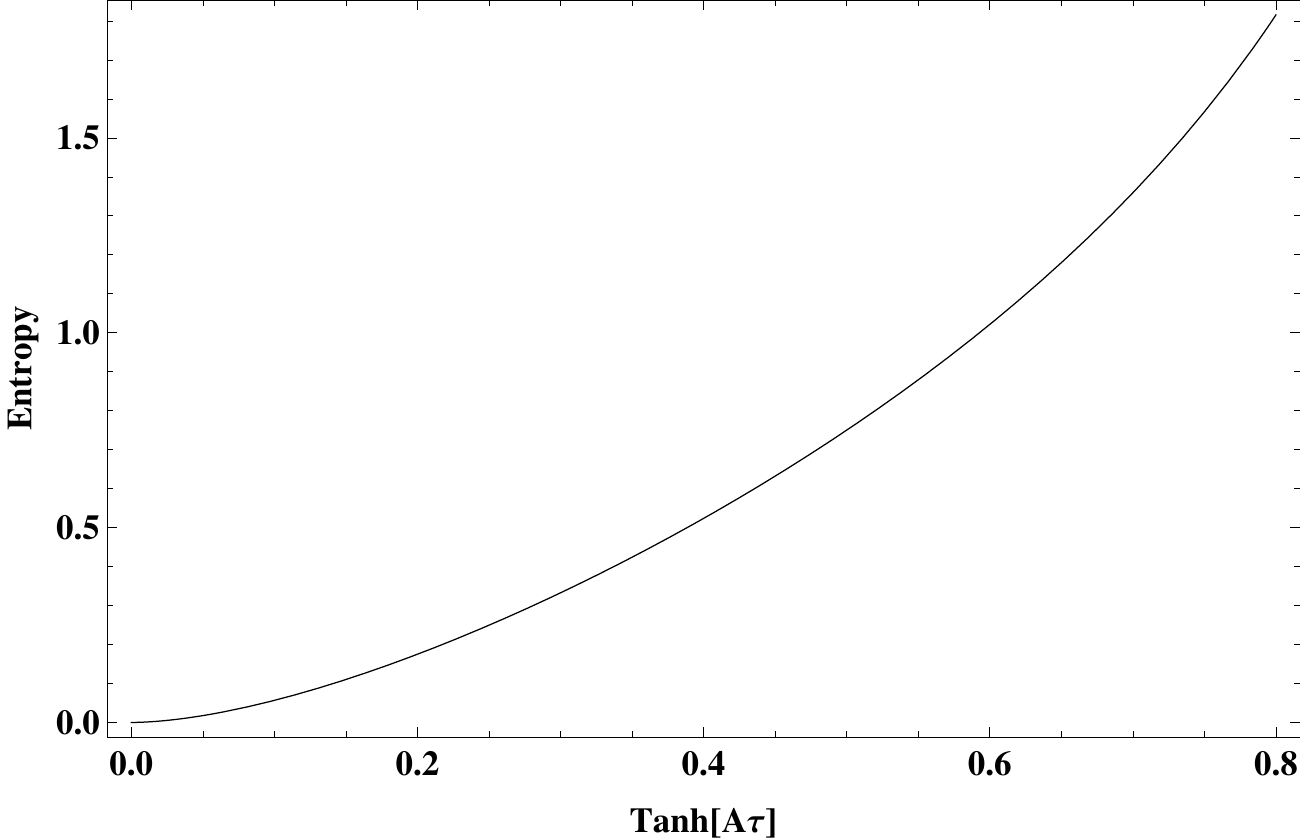}
\caption{Entropy of the outgoing modes  corresponding to the case of  a classical  black-hole.}\label{entropysu11}
\end{figure}
\end{center}

\begin{center}
\begin{figure}
\includegraphics[height=3in,width=4.8in]{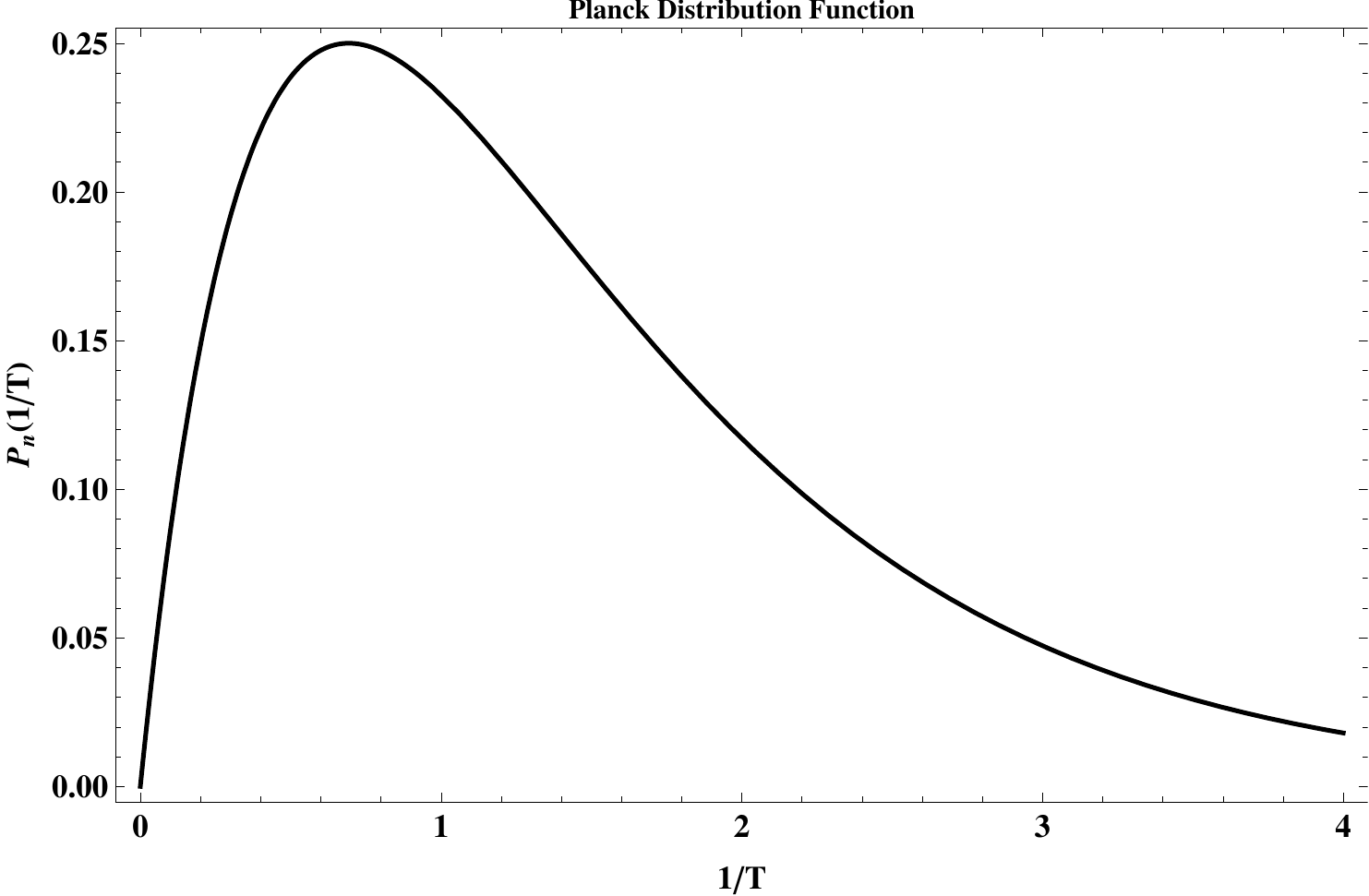}
\caption{Planckian distribution of the outgoing radiation }\label{plank}
\end{figure}
\end{center}

The general  Hamiltonian  in the model system of interaction between the particle and black hole modes is given  in the trilinear form 
\begin{equation} \label{trl-ham0}
\mathcal{H} = \omega_a \ a^\dag a + \omega_b \ b^\dag b + \omega_c \ c^\dag c + \kappa \ a b^\dag c^\dag + \kappa^\ast \ a^\dag b c \ .
\end{equation}
For ease of calculation and comparison with the results of \cite{nb},  we define $\kappa=i\hbar\chi$ to get 

\be \label{trl-ham1}
\mathcal{H} = \omega_a \ a^\dag a + \omega_b \ b^\dag b + \omega_c \ c^\dag c
+ i \hbar \chi a b^\dag c^\dag - i \hbar \chi^{*} a^\dag b c.
\ee
Here $a$, $b$, and $c$ are the boson annihilation operators of the quantized black hole modes, the outgoing modes and the incoming radiation modes respectively. Energy conservation requires $\omega_a = \omega_b + \omega_c$. In order to show the entanglement between the modes $b$ and $c$, we first treat the black hole mode  $a$ as classical (c-number $A$) to give the Hamiltonian
\be \label{bi-ham1}
\mathcal{H} =  \omega_b \ b^\dag b + \omega_c \ c^\dag c
+ i \hbar\chi A b^\dag c^\dag -i \hbar\chi^{*} A b c \ .
\ee
This Hamiltonian can be written in terms of the generators of an $su(1,1)$ algebra
\begin{equation}
K_+ = b^\dag c^\dag \ , \qquad K_- = b c \ , \qquad K_0 = \frac{(b^\dag b + c^\dag c + 1)}{2},
\end{equation}
with  $\omega_b = \omega_c = \omega$, as
\be \label{bi-ham2}
\mathcal{H} =  \omega(2 K_0-1) + i \hbar \chi A K_+ -i \hbar \chi^{*}A K_-\ .
\ee
The corresponding Casimir is
\be \label{scriptK}
\mathcal{K} = K_0^2 - \frac{K_+ K_- + K_- K_+}{2}=\frac{(c^{\dag}c - b^{\dag}b)^2 -1}{4}.
\ee
The representations are labelled by the eigenvalues of the Casimir operator and the number operator for the outgoing modes$n_b$:
\be
\mathcal{K} \, \big\vert k,n_b  \big\rangle   =  k(k-1) \, \big\vert k, n_b  \big\rangle  \
\ee
The action of the su(1,1) generators on the basis $\big\vert k, n_b  \big\rangle$ is
\begin{subequations}
\begin{align}
K_0 \, \big\vert k, n_b  \big\rangle  &=   (k + n_b) \, \big\vert k, n_b \big\rangle \ , \\
K_+ \, \big\vert k, n_b  \big\rangle  &=    \sqrt{(n_b+1)(n_b+2k)} \, \big\vert k, n_b+1  \big\rangle \ , \\
K_- \, \big\vert k, n_b  \big\rangle  &=    \sqrt{n_b(n_b + 2k -1)} \, \big\vert k, n_b-1  \big\rangle \ 
\end{align}
\end{subequations}
A convenient basis to use is the two-mode oscillator basis
\be
\big\vert n_b, n_c  \big\rangle=\frac{b^{ \dag n_b } c^ {\dag n_c}}{n_b! \ n_c!} \ \big\vert 0,0 \big\rangle \ .
\ee
 We set $n_c-n_b=q$, with $q$ being called the degeneracy parameter and takes values $q = 0,1,\cdots$. It is easy to go from one basis to another by noting $k=\frac{1}{2}(1+q)$. 
Since the Hamiltonian is linear in the generators of the  Lie algebra of the $SU(1,1)$ group, the unitary evolution due to the interaction is given by
$\vert\Psi(t)\rangle$,  the  coherent state relevant to the discrete series representation of $SU(1,1).$
\begin{equation}
\big\vert \Psi(t) \big\rangle =\exp  \big[\beta(t) K_+ -\beta^*(t) K_-\big] \big\vert k,0 \big\rangle \ .
\end{equation}
 Where $\beta(t) = \chi A t$ and the kinetic part contributes  an overall  phase factor. Using the disentanglement theorem (Zassenhaus formula) for su(1,1) we get 
\begin{equation}
\exp  \big[\beta(t) K_+ -\beta^*(t) K_-\big ]=\exp \big[\gamma ' K_-\big] \exp \big[\eta K_0\big]\exp \big[\gamma K_+\big],\label{df}\end{equation}
where, $\gamma = \tanh (|\beta(t)|)e^{i\phi}$, $\eta=2 \ln \cosh(|\beta|)$, and $\gamma'=-\gamma^{*}$.
This allows us to write the time evolved wave function as:
\begin{equation}
\big\vert \Psi(t) \big\rangle = \big(1 - \gamma^2 \big)^k \sum_{n_b=0}^\infty \sqrt{\frac{\Gamma(2k+n_b)}{n_b! \ \Gamma(2k)}} \ \gamma^{n_b} \  \big\vert k, n_b \big\rangle \ ,
\end{equation}
In the  two mode oscillator basis this  becomes:
\begin{equation}
\big\vert \Psi(t) \big\rangle = \big(1 - \gamma^2 \big)^{(q+1)/2} \sum_{n_b=0}^\infty \sqrt{\frac{(n_b+q)!}{n_b! \ q!}} \ \gamma^{n_b} \big\vert n_b,n_b+q \big\rangle \ .
\end{equation}
 For the special case $q=0$ (overall neutrality),
\begin{equation}
\big\vert \Psi(t) \big\rangle = \sqrt{1 - \gamma^2} \sum_{n_b=0}^\infty \ \gamma^{n_b} \ \big\vert n_b,n_b \big\rangle \ .
\end{equation}
The density matrix is
\be
\rho(t)=\sum_{n_b,m_b}C_{n_b,n_b+q}\ C^\ast_{m_b+q,m_b}\ \big|n_b,n_b+q \big\rangle \big\langle m_b+q,m_b \big| \ \label{rho} ,
\ee
where
\be
C_{n_b,n_b+q} = \big(1 - \gamma^2 \big)^{(q+1)/2}  \ \sqrt{\frac{(n_b+q)!}{n_b! \ q!}} \ \gamma^{n_b} \ .
\ee
The entanglement properties are calculated by taking the  partial transpose of $\rho$ :
\be
\rho(t)=\sum_{n_b,m_b}C_{n_b,m_b+q}\ C^\ast_{n_b+q,m_b}\ \big|n_b,m_b+q \big\rangle \big\langle n_b+q,m_b \big| \ .
\ee
When  $n_b=n_c=n$ ,$(q=0)$,
\be
\rho_{PT}(t)=\sum_{n,m}C_{n,m}\ C^\ast_{m,n}\ \big|n,m \big\rangle \big\langle m,n \big|
\ee
Where $C_{n,m}=\tanh^{m+n}(\chi A t) \mathrm{sech}(\chi A t)$. The Eigenvalues can be found by   looking at the terms of the matrix \cite{agarwal}
\begin{equation}
C_{n,n} C^\ast_{m,m}|n,m \rangle \langle m,n| + C_{m,m} C^\ast_{n,n}|m,n\rangle \langle n,m|,
\end{equation}
and are
\bea
\lambda_{nn}=\tanh^{2n}(\chi A t) \mathrm{sech}^2(\chi A t); \hspace{1in}  n=m   \\
\lambda_{nm}=\pm\tanh^{m+n}(\chi A t) \mathrm{sech}^2(\chi A t);\hspace{1in} n\ne m
\eea
Since the eigenvalues can be negative ,  the state is entangled, by the Peres-Horodecki citerion. 
The probability distribution of the outgoing modes is given by
$P_{n_b} = sech^2(A\tau) \, \tanh^{2n_b}(A\tau)$
and a plot of $P_{n_b}$  vs. $n_b$ for different values of the scaled time $\tau$ is shown fig. (\ref{photondist_su11}).
The Entropy of entanglement is given by
\bea
S=-\mathrm{Tr}_{n}\rho \log \rho = \cosh^2(\chi A t) \log \cosh^2(\chi A t) - \log \sinh^2(\chi A t)
\eea
Figure (\ref{entropysu11}) shows how the entanglement entropy increases with the scaled time $\tau$ and it is clear that as $\tau \longrightarrow \infty$ the states $b$ and $c$ are maximally entangled.
To show the thermal nature of the outgoing radiation, we define the temperature 
$T=[\hbar \omega_c/2 k_B \ln(\mathrm{coth}(A \tau)]$ allows us to write the entropy in a more familiar form:
\be
S=-\log(1-e^{\frac{h\omega}{k_BT}})-\frac{h\omega}{k_BT}(1-e^{\frac{h\omega}{k_BT}})^{-1}
\ee
The probabilty distribution  of the outgoing modes, $P_{n_b}(1/T)=e^{-2/T}[e^{1/T}-1]$ is plotted in fig. (\ref{plank}) as a function of $1/T$,  showing that  the outgoing modes have a thermal distribution in the form predicted by Hawking for black holes.
We therefore see that in the long time limit there is  maximal entanglement between the incoming and outgoing modes. We also see that a thermal distribution  results for the outgoing radiation, as expected, when the black hole is considered to be a classical object.
In the next section we quantize the black hole modes and reexamine the thermal and entanglement properties.
\section{Quantized Black-hole}

We now come to the advantage that analogue models provide. In the absence of a quantum theory of gravity, it is only through models such as the one we are studying, that we can actually examine the result of quantizing the ``black hole'' modes.
Such effects have only  been studied   in the short time limit \cite {nb}, which clearly is not sufficient to determine effects at infinity as required by black hole physics.
 We would like to see if the thermal properties and the entanglement properties in the long time limit change when the black hole is quantized.
In this section, we shall show how quadratic algebras can be used to solve this particular problem. 

The full Hamiltonian with three quantized modes can be written as in eqn.(\ref{trl-ham0}),
\be
\mathcal{H} = \omega_a \ (a^\dag a + K_0) + i \hbar\chi a K_+ - i \hbar\chi^* a^\dag K_- \ ,
\label{trl-ham2}
\ee
by choosing $\omega_b=\omega_c=\omega_a/2$.

Defining
\be
Q_0 = \frac{1}{2}(K_0 - a^{\dag}), \qquad Q_+ = a K_+, \qquad Q_- = a^\dag K_-, \qquad \mathcal{L} = \frac{1}{2}(K_0 + a^\dag a) \ ,
\ee
the Hamiltonian Eq. (\ref{trl-ham2})  takes the form
\be
\mathcal{H} = 2 \omega_a \ \mathcal{L} + i \hbar\chi Q_+ - i \hbar\chi^* Q_- \ .
\label{trl-ham3}\ee
These operators  $Q_+,Q_-$ and $Q_0$ satisfy a \emph{quadratic algebra}, which is a special case of  {\it Polynomially deformed algebras}, introduced first by Sklyanin  \cite{sky}, in the context of braid groups  and modified to be used in quantum optics by Karassiov and Klimov \cite{kk}.
This quadratic algebra is given by,
\begin{align}
[Q_0,Q_\pm] &= \pm Q_\pm \no \\ \label{quad-alg1}
[Q_+,Q_-] &= 3 \ Q_0^2 + (2 \mathcal{L} -1) \ Q_0 - (\mathcal{K} + \mathcal{L}(\mathcal{L}+1)),
\end{align}
where, $\mathcal{L}$ is the central element of the quadratic algebra and $\mathcal{K}$, defined in equation(\ref{scriptK}), is the Casimir for the $su(1,1)$ algebra. The Casimir Operator for this algebra is
\be \label{cas-quad1}
\mathcal{C} = Q_+ Q_- + Q_0^3 + (\mathcal{L}-2)\ Q_0^2 - (\mathcal{K} + \mathcal{L}^2+2\mathcal{L}-1)\ Q_0 + [\mathcal{K} + \mathcal{L}(\mathcal{L}+1)].
\ee
This is a special form of the  generic  $su(1,1)$ quadratically deformed polynomial algebra
\be \label{quad-alg2}
[Q_+,Q_-] = \alpha \ Q_0^2 + \beta \ Q_0 + \zeta  \ \equiv g(Q_0) - g(Q_0-1)\ \equiv  f(Q_0)
\ee
where the function, $g$ is called the structure function and is defined upto the addition of a constant as,
\be \label{sf1}
g(Q_0) = \frac{\alpha}{3} Q_0^3 + \frac{(\alpha+\beta)}{2} Q_0^2 + \frac{(\alpha+ 3 \beta+6\zeta)}{6} Q_0 \ .
\ee
The Casimir expressed in terms of the structure function is
\be
\mathcal{C} = Q_- Q_+ + g(Q_0) = Q_+ Q_- + g(Q_0 -1) \ .
\ee
The quadratic algebra (\ref{quad-alg1})  corresponds to the choice  $\alpha=3, \beta=(\mathcal{L}-1)$, and $\zeta=-[\mathcal{K}+\mathcal{L}(\mathcal{L}+1)]$. The quadratic algebra belongs to a class of algebras similar to the enveloping algebras of $sl(2,C)$ and has  a notion of highest weight modules and every finite dimensional module is semi-simple \cite{smith}.

Thus, the basis $\big\vert k, \ell, n_b  \big\rangle $ of a unitary representation of the quadratic algebra  is finite dimensional and characterized by the quantum numbers $k$ and $\ell$ \cite{vsk-bb1}. The action of the generators is given by
\begin{subequations}
\begin{align}
\mathcal{L} \, \big\vert k, \ell, n_b  \big\rangle  &=   \ell \, \big\vert k, \ell, n_b  \big\rangle \ , \label{quadrepcent}\\
Q_0 \, \big\vert k, \ell, n_b  \big\rangle  &=   (k + n_b -\ell) \, \big\vert k, \ell, n_b  \big\rangle \ , \label{quadrep0} \\
Q_+ \, \big\vert k, \ell, n_b  \big\rangle  &=    \sqrt{(n_b+1)(n_b+2k)(2\ell-k-n_b)} \, \big\vert k, \ell, n_b + 1  \big\rangle \ , \label{quadrepp} \\
Q_- \, \big\vert k, \ell, n_b  \big\rangle  &=    \sqrt{n_b(n_b+2k-1)(2\ell-k-n_b+1)} \, \big\vert k, \ell, n_b - 1 \big\rangle \ .\label{quadrepm}
\end{align}
\end{subequations}
With $n_b = 0, 1, 2, \cdots, (2\ell-k)$.

For our purposes it is better to redefine the states in terms of the following constants, $n_c - n_b = q = 2k-1$ ,  $p=n_c+n_a$ and $ p-q=2 \ell-k$. With the new states, the action of the generators can be written down as follows:
\begin{subequations}
\begin{align}
Q_0 \, \big\vert p-q-n_b, n_b+q, n_b \big\rangle  &=   \left( \frac{4 n_b+3q-2p+1}{4} \right) \, \big\vert p-q-n_b, n_b+q, n_b \big\rangle \ , \\
Q_+ \, \big\vert p-q-n_b, n_b+q, n_b \big\rangle  &=    \sqrt{(p-q-n_b)(n_b+q+1)(n_b+1)} \, \big\vert p-q-n_b-1, n_b+q+1, n_b+1 \big\rangle \ ,  \\
Q_- \, \big\vert p-q-n_b, n_b+q, n_b \big\rangle  &=    \sqrt{(p-q-n_b+1)(n_b+q)n_b} \, \big\vert p-q-n_b+1, n_b+q-1, n_b-1 \big\rangle \ .
\end{align}
\end{subequations}
From the structure of the Hamiltonian, it is evident that the incoming and outgoing particle modes are being created through the annihilation of the black hole modes. Therefore we have the following conditions imposed by particle number conservation: $n_c=n_b$, $n_a=\frac{n_c+n_b}{2}$, hence, $q=0$ and $p=2n_b$.
We start with a state in which all the energy modes are concentrated in the black hole. Thus,
\begin{subequations}
\begin{align}
Q_0 \, \big\vert p-n_b, n_b, n_b \big\rangle  &=   \left( \frac{4n_b-2p+1}{4} \right) \, \big\vert p-n_b, n_b, n_b \big\rangle \ , \\ \label{qplus}
Q_+ \, \big\vert p-n_b, n_b, n_b \big\rangle  &=    \sqrt{(p-n_b) (n_b+1)(n_b+1)} \, \big\vert p-n_b-1, n_b+1, n_b+1 \big\rangle \ ,  \\
Q_- \, \big\vert p-n_b, n_b, n_b \big\rangle  &=    \sqrt{(p-n_b+1) n_b \ n_b} \, \big\vert p-n_b+1, n_b-1, n_b-1 \big\rangle \ .
\end{align}
\end{subequations}

With the inclusion of the quantized black hole modes, the Hamiltonian can be written as a linear combination of the generators of the quadratic algebra. Since $\mathcal{L}$ is the central element of the quadratic
algebra , the time evolved state of  the Hamiltonian is
 the Perelemov coherent state (PCS) of the quadratic algebra:
\be \label{csone}
\big\vert \Psi(\tau)
\big\rangle =  \exp \tau(Q_+ - Q_-) \big\vert p, 0, 0 \big\rangle \, \ee
where $\tau$ is the dimensionless time $\chi t$.
The ground state deserves some comments. The initial state is not completely devoid of energy, this is indicated by the degeneracy factor $p \neq 0$, in other words, in the initial state all the energy is concentrated in the BH.

As it stands the exponential can not be disentangled due to the nonlinear terms occurring in the commutation relation. 
 We   have used deformed algebra techniques to develop a Zassenhaus formula for quadratic algebras to calculate the coherent states in \cite{sbpv}. For the sake of completeness, we will discuss this construction in some detail.

The first step is to map the  quadratic algebra to $su(1,1)$ or $su(2)$ as the case may be, this process is called the linearization of the algebra. This technique relies on the crucial step of constructing a new annihilation operator $\wt{Q}_-$ such that it satisfies the following commutation relations $[Q_+,\wt{Q}_-]= - 2 \lambda Q_0$ and $[Q_0,\wt{Q}_-] = -2\wt{Q}_-$. When $\lambda =\pm 1$ the algebra is either $su(1,1)$ or $su(2) $respectively. To find this new operator $\wt{Q}_-$ we make an ansatz 
\be
\wt{Q}_- = F(\mathcal{C},Q_0) \ Q_- ,
\ee
where $F$ is chosen to be a function of the Casimir and the diagonal generator. This function can be solved for by substituting this ansatz in the linearized commutator, i.e.,
\bea \no
Q_+ \ F(\mathcal{C},Q_0) \ Q_- - F(\mathcal{C},Q_0) \ Q_- \ Q_+ & = & - 2 \lambda Q_0 \\
F(\mathcal{C},Q_0-1) \ Q_+ \ Q_- - F(\mathcal{C},Q_0) \ Q_- \ Q_+ & = & - 2 \lambda Q_0 \ .
\eea

In the present case it is
\be
F(\mathcal{C},Q_0) = \frac{Q_0(Q_0+1) \lambda + \epsilon}{\mathcal{C} - g(Q_0)}.
\ee
Where $\epsilon$ is an arbitrary constant that can be fixed by noting that the ground state of $Q_-$ and $\wt{Q}_-$ is same. Note that in the present study $\lambda = 1$. Details of this construction can be found in \cite{sbpv}.

Now that we have a new lowering operator, the PCS is easy to construct and is given by
\be \label{cstwo}
\big\vert \Psi(\tau) \big\rangle =  \exp \tau (Q_+ - \wt{Q}_-) \big\vert p, 0, 0 \big\rangle \ .
\ee
 Note that Eq. (\ref{csone}) is equivalent to Eq. (\ref{cstwo}), since in both cases the annihilation operator acting on the ground state gives zero and the diagonal operator gives trivial phase and can be absorbed into the normalization constant. Thus the operator responsible for the time evolution of the state is $Q_+$. Having said that, one must bear in mind that the evolution given by Eq. (\ref{cstwo}) is exact and valid for even large times unlike the evolution advocated by \cite{nb}.

 Since the algebra is linearized we can apply  the disentanglement formula  (\ref{df}) to Eq. (\ref{cstwo}) to get
\be
\big\vert \Psi,\gamma(\tau )\big\rangle = \frac{1}{\sqrt{N(\tau)}} \ \exp(\gamma \ Q_+) \big\vert p, 0, 0 \big\rangle \ ,
\ee
where $N(\tau)$ is a normalization constant  and  $\gamma = \tanh(\tau) ;\;\;  \tau=\chi t$. Using Eq. (\ref{qplus}), the CS is 
\be \label{cs-trl1}
\big\vert \Psi,\gamma(\tau) \big\rangle = \frac{1}{\sqrt{N(\tau)}} \sum_{n_b=0}^p \sqrt{\frac{\Gamma(p+1)}{\Gamma(p-n_b+1)}} \ \gamma^{n_b} \ \big\vert p-n_b, n_b, n_c \big\rangle \ ,
\ee
and the normalization constant is
\be
N(\tau) = \exp[\coth (\tau)^2] \ \tanh^{2p} (\tau) \ \Gamma\big(p+1,\coth (\tau)^2 \big) \ .
\ee
$\Gamma(a,b)$ is the reduced gamma function. For the sake of completeness the above CS in the $\big\vert k, \ell, n_b \big\rangle$ basis is
\be \label{cs-trl2}
\big\vert \Psi,\gamma(\tau) \big\rangle = \frac{1}{\sqrt{N(\tau)}} \sum_{n_b=0}^{2\ell-k} \ \sqrt{\frac{\Gamma(2k+n_b) \Gamma(2\ell-k+1)}{\Gamma(2k) \Gamma(n_b+1) \Gamma(2\ell-k-n_b+1)}} \ \gamma^{n_b} \ \big\vert k, \ell, n_b \big\rangle \ .
\ee
The normalization constant for the above CS is given by the Tricomi confluent hypergeometric function $U(a,b,x)$,
\be
N(\tau) = \gamma^{-4k} \ U \big(2k,2\ell+k+1,\gamma^{-2} \big) \ .
\ee


The state of the system consisting of black hole (a)+incoming modes (c) + outgoingmodes (b) is a pure state and therefore the total entropy is zero.
Hence the marginal entropies of the black hole  and the particle-antiparticle
subsystems are equal:
$S_a = S_{bc}.$
From the Araki-Lieb \cite{araki} theorem, 
$|S_b -S_c| \ge S_{bc} = S_a\ge| S_b + S_c|. $
A quantitative measure of the entanglement between two subsystems is the index of correlation
$I_{x-y} = S_x + S_y- S_{xy}.$
In our case the index of correlation $I_{a-bc} $between the BH  and Particle- antiparticle  subsystems
is equal to twice the marginal entropy of Black Hole:$ I_{a-bc} = 2S_a.$

The quantum nature of the dynamics induced by the trilinear boson Hamiltonian leads
to strong entanglement between the black hole and the particle-antiparticle subsystems. This can be seen by calculating the reduced density operators of the subsystems.

\begin{center}
\begin{figure}
\includegraphics[height=3in,width=4in]{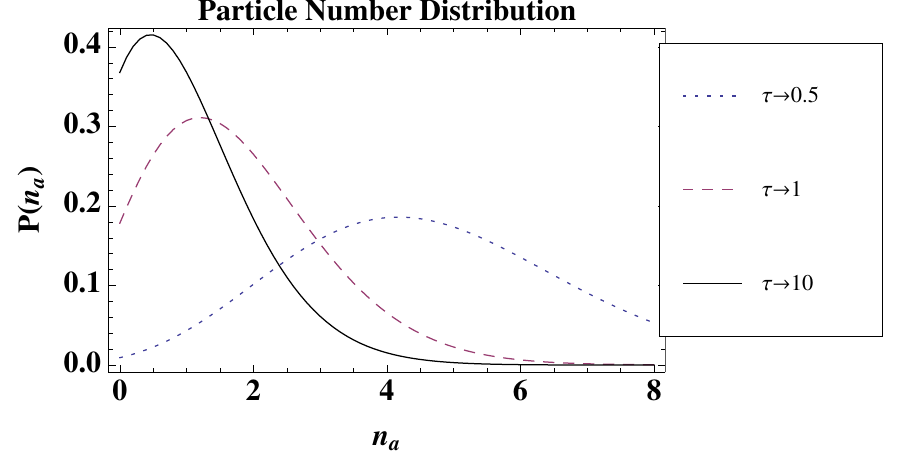}
\caption{ The particle number distribution for the  quantized black hole modes for different scaled times $\tau=\chi t.$}\label{fig3:trlpnda}
\end{figure}
\end{center}

\begin{center}
\begin{figure}
\includegraphics[height=3in,width=4in]{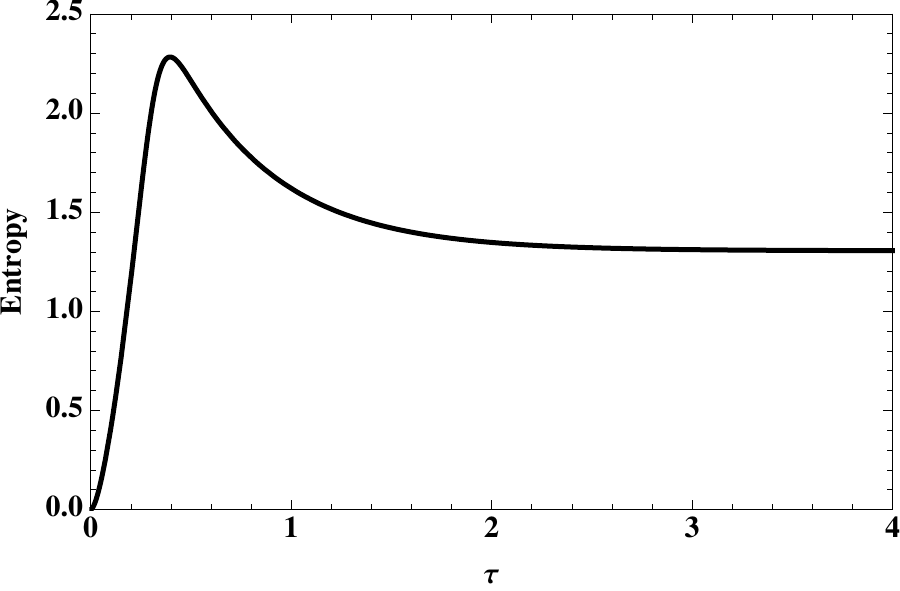}
\caption{Entropy for the black hole modes with  scaled time $\tau=\chi t.$}\label{fig5:trlentropya}
\end{figure}
\end{center}

The time dependent  density operator corresponding to the system  Eq. (\ref{cs-trl1}) is,
\be
\rho_{abc}(\tau) = \frac{1}{N(\tau)} \sum^{p}_{n_b,m_b} \frac{\Gamma(p+1) \gamma^{n_b+m_b}(\tau)\big\vert p-n_b,n_b,n_b\big\rangle \big\langle m_b,m_b,p-m_b\big\vert}{\sqrt{\Gamma(p-n_b+1) \Gamma(p-m_b+1)}}  .
\ee
In the present case we have much richer structure as compared to the normal $su(1,1)$ case given in Eq.(\ref{rho}). The reduced density operator of the black-hole , obtained by  tracing  over the particle-antiparticle  modes, is ,
\be
\rho_{a} = \frac{1}{N(\tau)} \ \sum^{p}_{n_b=0} \ \frac{\Gamma(p+1)}{\Gamma(p-n_b+1)} \ \gamma^{2n_b} \ \big\vert p-n_b\big\rangle \big\langle p-n_b\big\vert \ .
\ee
%
%
%
Using the reduced density operator $\rho_a$, the particle number distribution (PND) for the BH modes is
\be
P(n_a) = \big\langle n_a \big\vert \rho_{a} \big\vert n_a \big\rangle = \frac{1}{N(\tau)} \ \frac{\Gamma(p+1)}{\Gamma(n_a+1)} \ \gamma^{2p-2n_a}  \ .
\ee
which is shown in  figure(\ref{fig3:trlpnda}). We see clearly that the long time behaviour is markedly different from the short time behaviour showing that studies utilising short time approximations would not see the behaviour exhibited here. The quadratic polynomial algebra methods we have introduced here allows us to see the full range of temporal behaviour. 
%
%
The entropy of the black hole is given by 
$S_a = - \sum_{n} P(n_a) \ \log[P(n_a)]$
and shown in figure(\ref{fig5:trlentropya}).
The positive nature of the entropy shows that the black hole modes are always entangled. In the evolution of the black hole modes, there is a rapid increase in its entropy as it interacts with the particle-antiparticle modes and it then settles down in the long time limit to an entangled state with lower entropy, having radiated away its available degrees of freedom. Further evidence for this is obtained below where we show the entropy of the outgoing modes.

\begin{center}
\begin{figure}
\includegraphics[height=2.5in,width=4in]{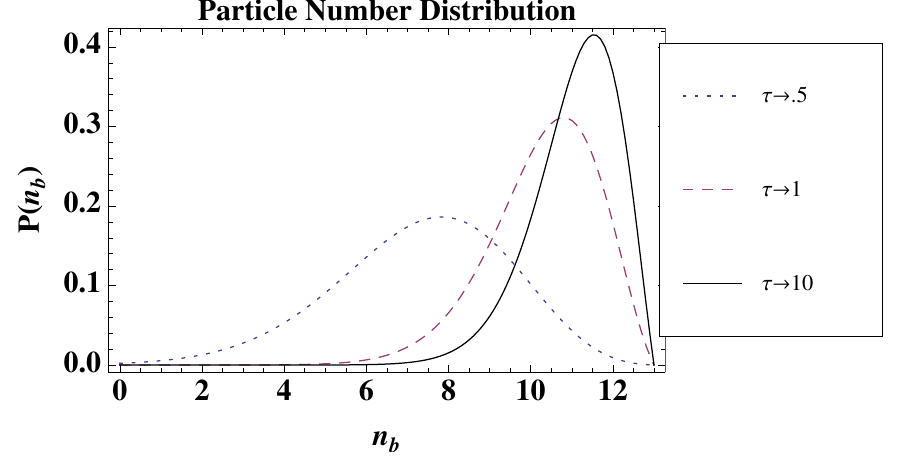}
\caption{Particle number distribution for the outgoing modes.}\label{fig6:trlpndb}
\end{figure}
\end{center}

The effective modification to the Hawking radiation with the incorporation of quantum entanglement with the black hole
modes can be found by studying various statistical properties of the outgoing modes ($n_b$). The reduced density matrix corresponding to the outgoing modes  obtained by summing over the black hole and incoming modes, is,
\be \rho_{b} =
\frac{1}{N(\tau)} \ \sum^{p}_{n_b=0} \ \frac{\Gamma(p+1)}{\Gamma(p-n_b+1)}
\ \gamma^{2n_b} \ \big\vert n_b\big\rangle \big\langle
n_b\big\vert \ .
\ee
The PND for the outgoing modes is given by 
\be
P(n_b) = \big\langle n_b \big\vert \rho_{b} \big\vert n_b \big\rangle = \frac{1}{N(\tau)} \ \frac{\Gamma(p+1)}{\Gamma(p-n_b+1)} \
\gamma^{2n_b}  \ 
\ee
and  is plotted in Fig. (\ref{fig6:trlpndb})
The plots for the  mean particle  numbers  for the black hole modes and the  outgoing modes are depicted in figures (\ref{fig4:trlmpna}) and  ({\ref{fig7:trlmpnb}).
\begin{center}
\begin{figure}
\includegraphics[height=2.5in,width=3.5in]{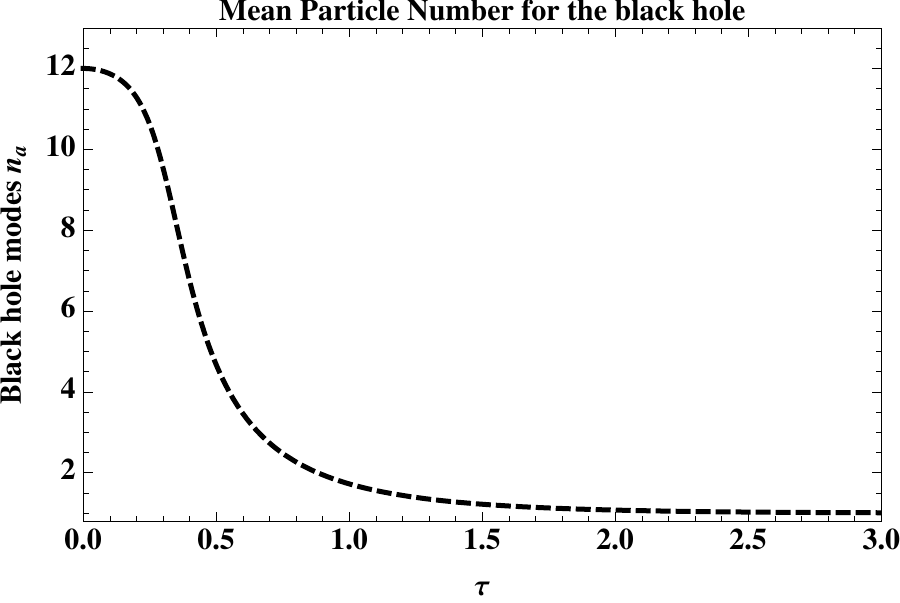}
\caption{Mean  particle number of the black hole modes as function of time.}\label{fig4:trlmpna}
\includegraphics[height=2.5in,width=3.5in]{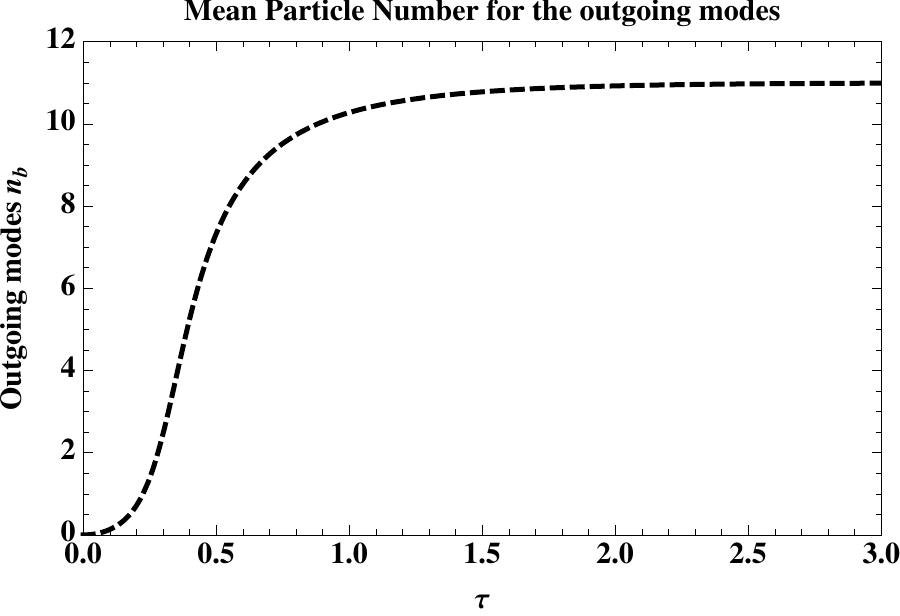}
\caption{Mean particle number  for the outgoing modes as a function of  time.}\label{fig7:trlmpnb}
\end{figure}
\end{center}
Finally the entropy for the outgoing modes is 
$S_b = - \sum_{n_b} P(n_b) \ \log[P(n_b)],$
and is shown  in Fig. (\ref{fig8:trlentropyb}). As in the case of the black hole entropy, we see again that the entropy of the outgoing modes also shows a limiting, asymptotic value of a finite non-zero entropy and suggests that there may be a final stable state of the black hole still entangled with the outgoing radiation detected at infinity. This clearly warrants an in depth study of the gravitational black hole with as much quantization that our existing techniques allow.
\begin{center}
\begin{figure}
\includegraphics[height=3in,width=4in]{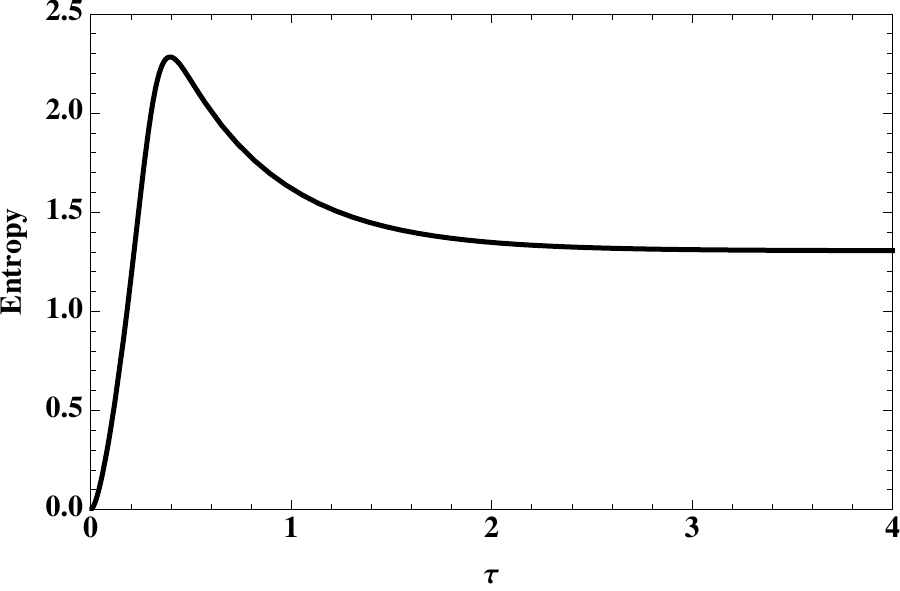}
\caption{Entropy for the outgoing modes as a function of time.}\label{fig8:trlentropyb}
\end{figure}
\end{center}
Let us compare the graphs of Figs. (\ref{fig3:trlpnda}) and (\ref{fig4:trlmpna}) i.e. PND and mean particle  number graphs corresponding to the BH modes with those of the outgoing modes shown in Figs. (\ref{fig6:trlpndb}) and (\ref{fig7:trlmpnb}) respectively. It is clear from the graphs that at the initial time all the energy is concentrated in the BH modes. As time increases particle-hole pairs are created. Those pairs that are created at the event horizon have very important role to play. In the pair that is created one particle goes into the BH (ingoing modes) and the other escapes the event horizon and can be detected, with a particle number distribution given by Fig. (\ref{fig6:trlpndb}) . Let us re-emphasize that these plots are physically relevant for long time only due to fact that we have identified a quadratic algebra structure of the model.  In similar fashion to the classical black hole case, we can introduce a temperature into the system , and compare the resulting distribution of the outgoing modes with the Planck distribution  shown  earlier in Fig.(\ref{plank}). 
\begin{center}
\begin{figure}
\includegraphics[height=3.5in,width=5.5in]{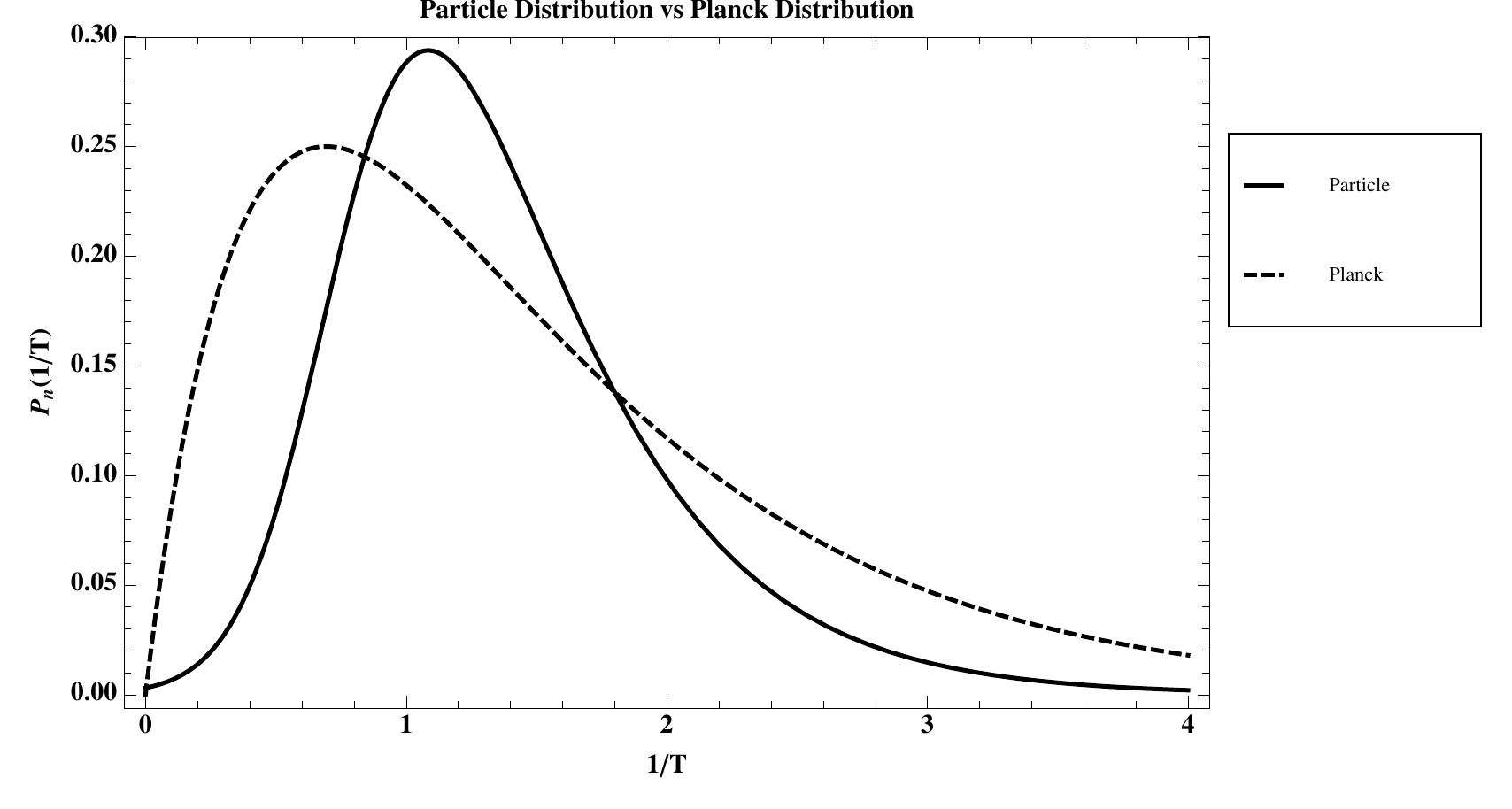}
\caption{Number Distribution for the outgoing modes as a function of temperature with quantized BH modes (solid line) for p=6 modes, and classical BH (dashed line) }\label{hawk1}
\end{figure}
\end{center}
\begin{center}
\begin{figure}
\includegraphics[height=3.5in,width=5.5in]{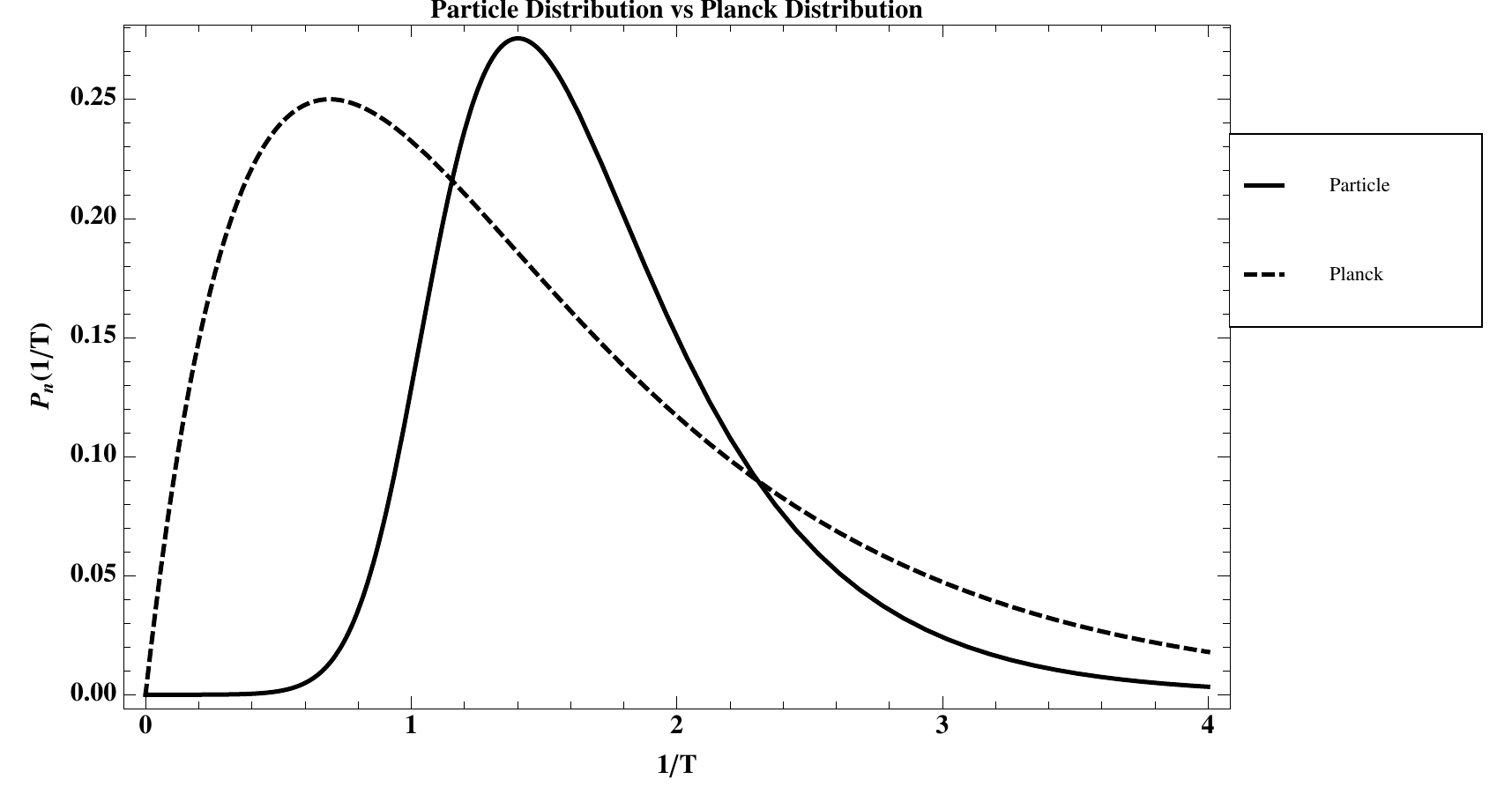}
\caption{Number Distribution for the outgoing modes as a function of temperature with quantized BH modes (solid line) for p=10 modes, and classical BH (dashed line) }\label{hawk2}
\end{figure}
\end{center}
From the two  figures (\ref{hawk1}) and (\ref{hawk2}) , we that there is a systematic shift from thermal behaviour of the outgoing radiation as the  number of quantized black hole modes are increased. This clearly illustrates the modification of the Planckian nature of Hawking radiation by the quantization of the black hole modes and shows that the outgoing radiation is ``superradiant''. To sum up, in this section we have considered the quantum optical analogue of hawking radiation from a quantized black hole as a parametric amplifier and used the polynomial deformation of a su(1,1) algebra. However, the structure of the Hamiltonian  is such that it can be viewed in a completely different manner, such that the underlying symmetry is a polynomially deformed su(2) algebra, which explains the physics of the ``superradiance'' in the system as will be shown in the next section.

\newpage
\section{Super Radiance in Black holes and the Dicke Model analogy}

In black hole physics, superradiance is the phenomenon of wave amplification through scattering off a rotating black hole and takes place when the wave frequency satisfies the superradiance condition: $ 0< \omega < m\omega_+$ where $\omega_+= \frac{J/M}{(r_H^2+(J/M)^2)} $, J is the angular momentum, M ,the mass and $r_H$, the horizon radius of the black hole.

Superradiance in quantum optics is the cooperative spontaneous emission  of photons from a collection of atoms. The concept of superradiance can be understood by picturing each atom as a tiny antenna emitting EM waves. Thermally excited atoms emit light with random intensity $\propto N$. Coherently excited atoms have EM field $\propto N$ and intensity $\propto N^2$ and leads to Dicke ''superradiance" \cite{dicke}.
In this section , we reframe the trilinear Hamiltonian model such that  the black hole modes and the incoming modes combine to form ''superradiant bound states'' which have an effective spin $\vec{J}$. This allows us to  populate the bound state (atomic) levels in  a similar way to that  of a rotating black hole with angular momentum.  We shall see that through the mapping of the Hamiltonian to that of the Dicke model,  the  system  allows for superradiance of the Dicke type.
By considering a rotating  Kerr Black hole as an excited state of the static Schwarzchild black hole,   Wald and Panangaden \cite{wald}  have already observed  that superradiance is analogous to  "Dicke superradiance " in atomic physics .

 Writing the trilinear Hamiltonian (\ref{trl-ham1}) in a form analogous to the   Dicke model (DM) in the rotating wave approximation\cite{nation}, with the incoming modes and black hole modes forming an atomic system (atomic modes) and the 
outgoing modes interacting with the atomic system, we get:
\be \label{dicke-ham1}
\mathcal{H}_{DM} = \omega \ (b^\dag b + J_0) + i \hbar \chi b^\dag J_- -i \hbar \chi^* b J_+ \ .
\ee
 $J_0, J_\pm$ form an  angular momentum algebra describing the atomic system formed by the black hole and incoming modes, which in the Schwinger representation can be written as
\begin{equation}
J_+ = a^\dag c \ , \qquad J_- = ac^\dagger \ , \qquad J_0 = \frac{(a^\dag a - c^\dag c)}{2} \ .
\end{equation}
 Thus, in this case, instead of the $b-c$ system in the earlier section,  it is the $a-c$ system  that acts like a single entity having angular momentum.

Defining the operators 
\be
P_0 = \frac{1}{2}(b^{\dag}b-J_0) , \quad P_+ = b^\dag \ J_-, \quad P_- = b \ J_+ , \quad \mathcal{L} = \frac{1}{2}(b^{\dag}b+J_0) \ .
\ee
the Hamiltonian (\ref{dicke-ham1})  becomes
\be \label{dicke-ham2}
\mathcal{H}_{DM} = 2 \omega \mathcal{L} + i \hbar \chi P_+ - i \hbar \chi^* P_-\ .
\ee
Now the algebra satisfied by $P's$ is 
\begin{align} \no
[P_0,P_\pm] & = \pm P_\pm , \\
[P_+,P_-] & = 3 P_0^2 - (2\mathcal{L}-1) P_0 - [\mathcal{J} + \mathcal{L}(\mathcal{L}+1)] \ .
\end{align}
Comparing with the earlier algebra in terms of the $Q's$ (\ref{quad-alg1}) , the linear term in $P_0$  is negative, and therefore the algebra is a polynomially deformed $su(2)$ algebra.
The basis states  of  the algebra are product states of the $su(2)$ algebra and the oscillator states:
\be \label{dicke-state1}
\big\vert j, \ell, m \big\rangle = \big\vert j, m \big\rangle  \ \big\vert n_b \big\rangle \ .
\ee
where $\vert n_b \rangle$ and $\{\vert j, m \rangle \}$ are the outgoing and $su(2)$ states respectively and $n_b=2l-m$. We construct a representation such that $\mathcal{J}$ and $\mathcal{L}$ are diagonal in the new combined basis
\be \label{diagbasis}
\mathcal{J} \ \big\vert j, \ell, m \big\rangle = j(j+1) \ \big\vert j, \ell, m \big\rangle \ , \qquad \mathcal{L} \ \big\vert j, \ell, m \big\rangle = \ell \ \big\vert j, \ell, m \big\rangle \ .
\ee
This can be divided into two different cases depending on the value $(2 \ell-j)$ takes. They are $(2 \ell-j) \geq 0$, $(2 \ell-j)< 0$. In both these scenarios the representation is that given below \cite{vsk-bb1}
\begin{subequations} \label{atomic-rep}
\begin{align}
P_0 \, \big\vert j, \ell, m  \big\rangle  &=   (\ell-m) \, \big\vert j, \ell, m  \big\rangle \ ,  \\
P_+ \, \big\vert j, \ell, m  \big\rangle  &=    \sqrt{(j+m)(j-m+1)(2 \ell-m+1)} \, \big\vert j, \ell, m + 1  \rangle \ ,  \\
P_- \, \big\vert j, \ell, m  \big\rangle  &=    \sqrt{(j-m)(j+m+1)(2 \ell-m)} \, \big\vert j, \ell, m - 1 \big\rangle \ .
\end{align}
\end{subequations}
We will now study the time evolution of the state of this model. Proceeding as already shown in the case of trilinear system we will construct a new operator $\wt{P}_-$ such that $[P_+, \wt{P}_-] = 2 P_0$. But for the initial state we find it convenient to use the representation wherein instead of $m$ we use $n_b$ making use of the central element. Therefore the representation for the algebra now acquires the form
\begin{subequations}
\begin{align}
P_0 \, \big\vert j, \ell, n_b  \big\rangle  &=   (n_b - \ell) \, \big\vert j, \ell, n_b  \big\rangle \ ,  \\
P_+ \, \big\vert j, \ell, n_b  \big\rangle  &=    \sqrt{(j+2 \ell-n_b)(j-2 \ell+n_b+1)(n_b+1)} \, \big\vert j, \ell, n_b + 1  \rangle \ ,  \\
P_- \, \big\vert j, \ell, n_b  \big\rangle  &=    \sqrt{(j-2 \ell+n_b)(j+2 \ell-n_b+1)(n_b)} \, \big\vert j, \ell, n_b - 1 \big\rangle \ .
\end{align}
\end{subequations}

Analogous to the previous case the time evolution of the state is a Perelomov Coherent state for $su(2)$
\be \label{dmcs1}
\big\vert \Phi(\tau) \big\rangle = \exp \tau (P_+ - \wt{P}_-) \big\vert j, \ell, 0 \big\rangle , 
\ee
where $\tau = \chi t$.  Using disentanglement formula for $su(2)$ we get 
\be
\big\vert \Phi(\tau) \big\rangle = \frac{1}{\sqrt{N(\tau)}} \ \exp(\gamma \ P_+) \big\vert j, \ell, 0 \big\rangle ,
\ee
where,   $\gamma = \tan(\tau)$.
The resultant time evolved state is :
\be \label{cs-dicke-field}
\big\vert \Phi(\tau) \big\rangle = \frac{1}{\sqrt{N(\tau)}} \sum_{n_b=0}^L \sqrt{\frac{\Gamma(2j-L+1+n_b) \Gamma(L+1)}{(n_b)! \Gamma(2j-L+1) \Gamma(L- n_b+1)}} \ \gamma^{n_b} \ \big\vert j, \ell, n_b \big\rangle \ .
\ee
where $L=2\ell+j$ ,
$N(\tau) = |\gamma|^{2L} \ \Psi\big(-L;-2j; 1/|\gamma|^2 \big) $  and
 $\Psi$ is the Tricomi confluent hypergeometric function.  The time evolved state can also be written in terms of the atomic modes as:
\be \label{cs-dicke-atom}
\big\vert \Phi(\tau) \big\rangle_{DM} = \frac{\hat{\gamma}^{j-L}}{\sqrt{N_{at}(\tau)}} \sum_{m=-j}^{L-j} \sqrt{\frac{\Gamma(j-m+1) \Gamma(L+1)}{\Gamma(L-j-m+1) \Gamma(2j-L+1) \Gamma(j+m+1)}} \ \hat{\gamma}^{m} \ \big\vert j, \ell, m \big\rangle \ .
\ee
where $\hat{\gamma} = \gamma^{-1}$ and $
N_{at}(\tau) = \frac{1}{|\hat{\gamma}|^{2L}} \frac{\Gamma(2j+1)}{\Gamma(2j-L+1)} {_1}F_1 \big(-L;-2j;|\hat{\gamma}|^2 \big)$. 
The density matrix for the state  Eq. (\ref{cs-dicke-field})  in terms of the field (outgoing) modes is given by
\be
\rho_{DM} = \big\vert \Phi(\tau) \big\rangle \big\langle \Phi(\tau) \big\vert = \frac{1}{N(\tau)} \sum_{n^\prime_b=0}^L \sum_{n^{\prime \prime}_b=0}^L \ C(n^\prime_b,L) C^\ast(n^{\prime \prime}_b,L) \ \big\vert j, 2\ell-n^\prime_b; n^\prime_b \big\rangle \big\langle n^{\prime \prime}_b; j, 2\ell-n^{\prime \prime}_b \big\vert \ .
\ee
For the sake of convenience we have abbreviated the summands  of Eq. (\ref{cs-dicke-field}) by $C(n^\prime_b,L)$.
This can also be written in terms of the atomic modes as 
\be
\rho_{DM} = \big\vert \Phi(\tau) \big\rangle \big\langle \Phi(\tau) \big\vert = \frac{\hat{\gamma}^{2(j-L)}}{N_{at}(\tau)} \sum_{m^\prime=-j}^{L-j} \sum_{m^{\prime \prime}=-j}^{L-j}  \ C(m^\prime,L) C^\ast(m^{\prime \prime},L) \ \big\vert j, m^\prime; 2\ell-m^\prime\big\rangle \big\langle 2\ell-m^{\prime \prime};j,m^{\prime \prime} \big\vert
\ee
The factors inside the summation in Eq. (\ref{cs-dicke-atom}) have been abbreviated by $C(m^\prime,L)$.
Performing a partial trace over the field modes we find the number distribution of the atomic modes as :
\be
P_{at} = \big\langle j,m \big\vert \rho_{at} \big\vert j,m \rangle  = \frac{|\hat{\gamma}|^{2(j-L)}}{N_{at}(\tau)} \frac{\Gamma(j-m+1) \Gamma(L+1)}{\Gamma(L-j-m+1) \Gamma(2j-L+1) \Gamma(j+m+1)} \ |\hat{\gamma}|^{2m} \,
\ee

This is plotted in Fig.(\ref{fig12:dickeatomdist})
\begin{center}
\begin{figure}
\includegraphics[height=3in,width=4.2in]{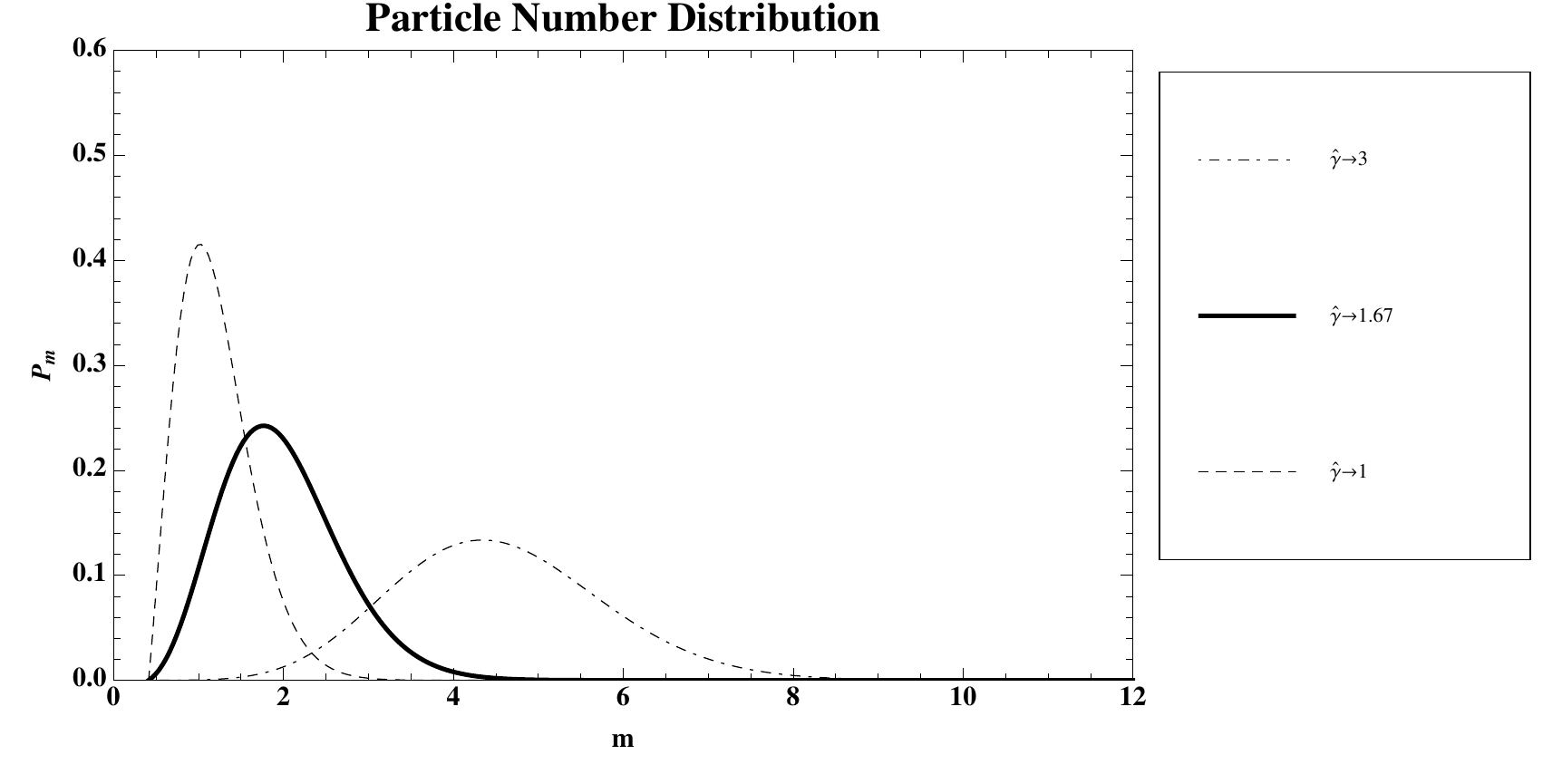}
\caption{ Particle number distribution of the atomic modes  for different times characterized by $\hat{\gamma}$.}\label{fig12:dickeatomdist}
\end{figure}
\end{center}

The entropy for the atomic modes  is
$S_{at} = - \sum_{m} \ P_{at} \log[P_{at}],$
and is plotted in Fig.(\ref{fig14:dickeentropyatom}).

\begin{center}
\begin{figure}
\includegraphics[height=3.2in,width=4.7in]{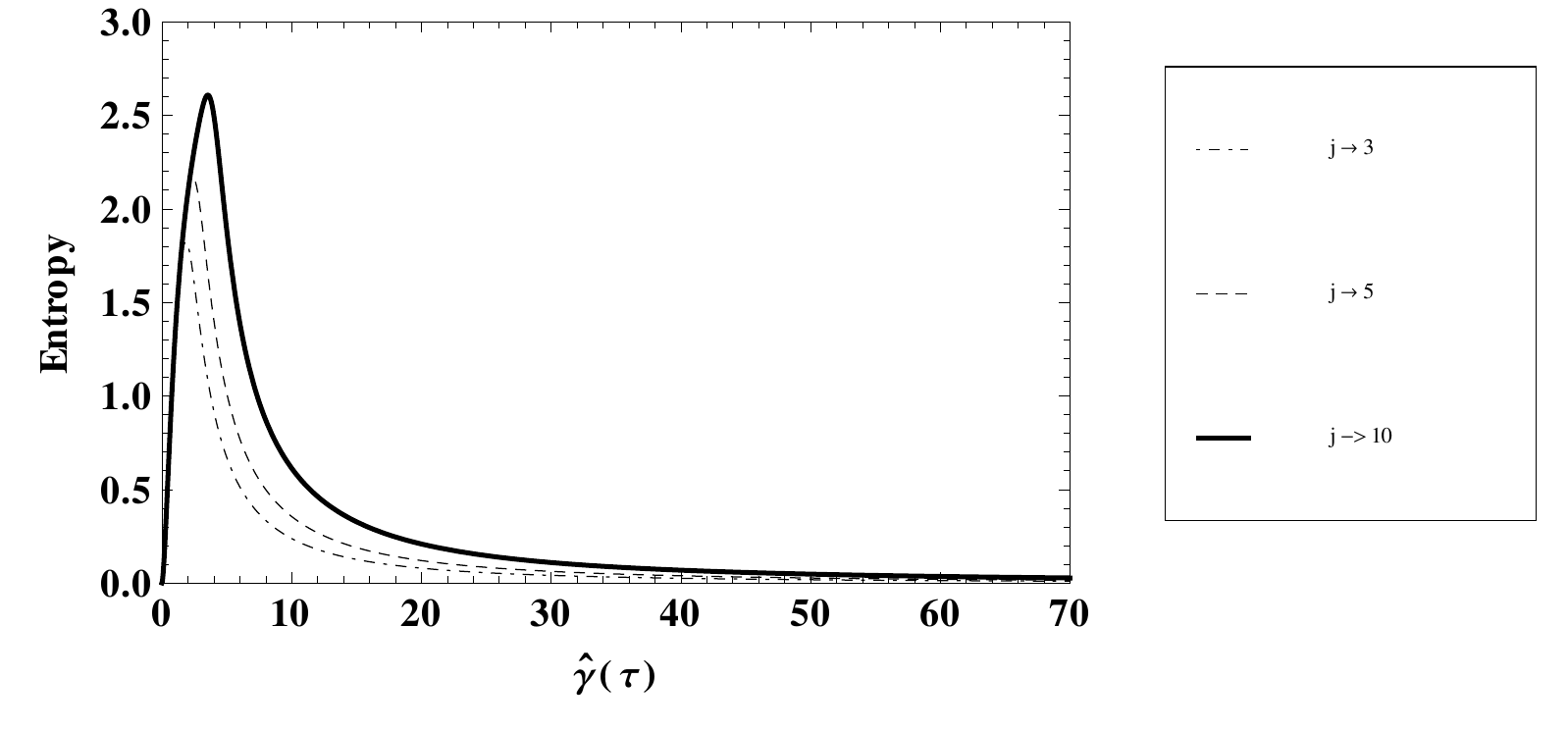}
\caption{Entropy of the atomic system for different number of atomic modes characterized by $j$.}\label{fig14:dickeentropyatom}
\end{figure}
\end{center}
From the plots we see that the distribution of the atomic modes becomes more symmetric with time and  as the number of modes increases, the entropy increases until it reaches a maximum value corresponding to a population of all the states and the system is maximally entangled. 
The particle number distribution for the field modes is
\be
P_{n_b} = \big\langle n_b \big\vert \rho_f \big\vert n_b \big\rangle = \frac{1}{N_f(\tau)} \frac{\Gamma(2j-L+1+n_b) \Gamma(L+1)}{\Gamma(n_b+1) \Gamma(2j-L+1) \Gamma(L- n_b+1)} \ |\gamma|^{2n_b}
\ee
with the normalization constant $N_f(\tau)$ being the same as that given in Eq. (\ref{cs-dicke-field}) and is plotted for different values of $\gamma$ in  Fig. (\ref{fig15:dickephotdist_b}).
\begin{center}
\begin{figure}
\includegraphics[height=3in,width=5.3in]{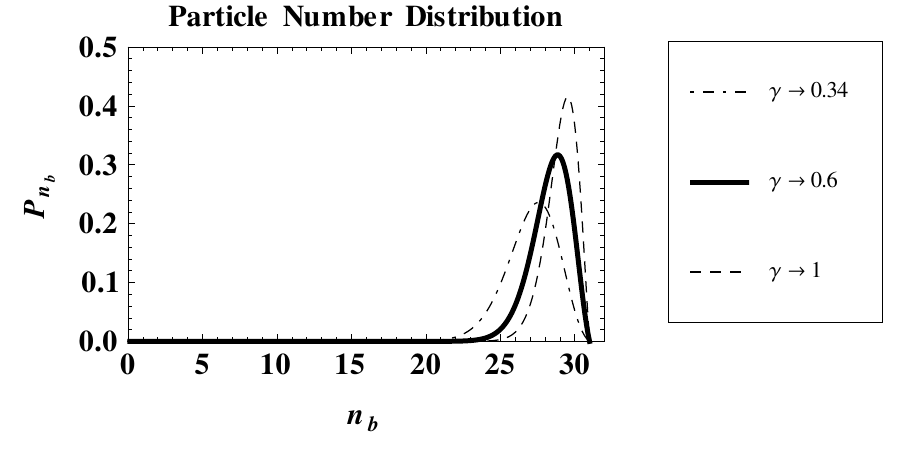}
\caption{Particle number distribution of the outgoing modes, for different times characterized by $\gamma$.}\label{fig15:dickephotdist_b}
\includegraphics[height=3in,width=5.1in]{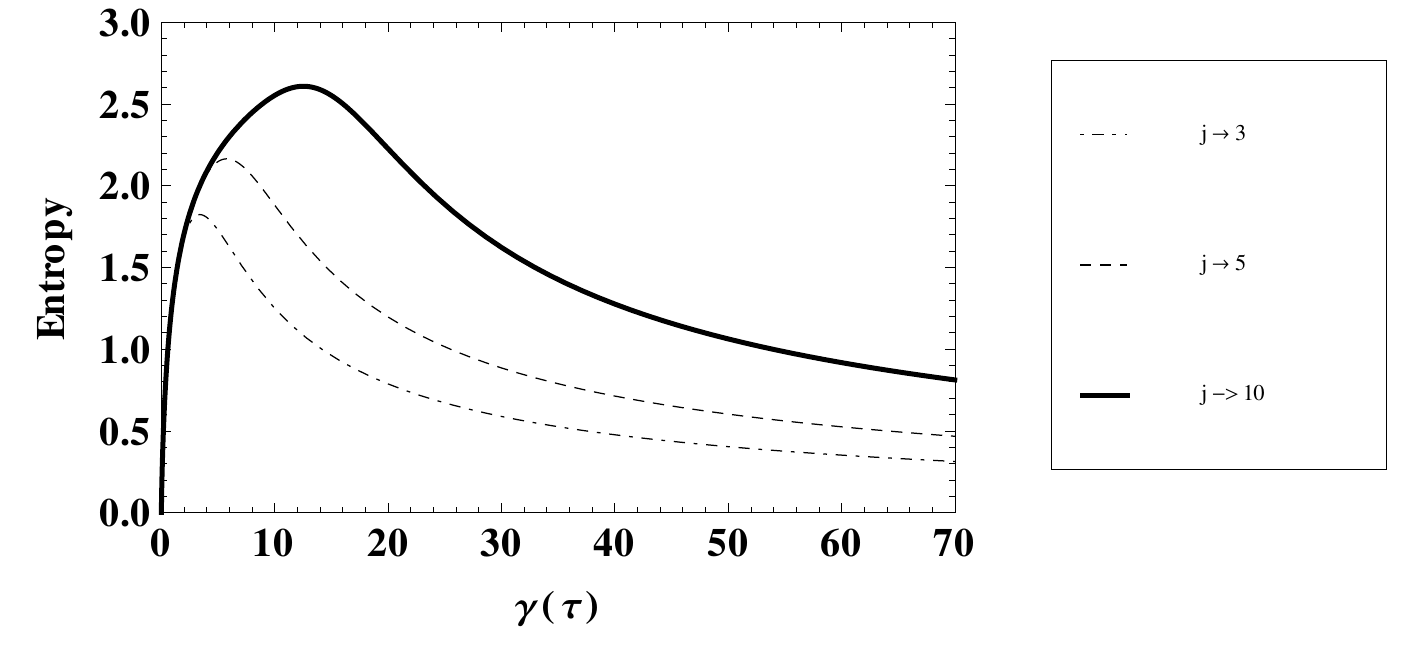}
\caption{Entropy of the outgoing modes, as a function of the time characterized by $\gamma$.}\label{fig16:dickeentropy_b}
\end{figure}
\end{center}
The entropy is given by
$S_b = - \sum_{n_b} \ P_{n_b} \log [P_{n_b}]$
and is plotted in Fig. (\ref{fig16:dickeentropy_b}).

From Fig. (\ref{fig15:dickephotdist_b}) , we see that the outgoing modes have a reciprocal behaviour to that of the atomic system, which is expected . The superradiant behaviour of the outgoing modes is evident when one compares  Fig. (\ref{fig15:dickephotdist_b}) with Fig. (\ref{photondist_su11}), and one sees clearly an enhancement in the outgoing radiation, which is due to the superradiant transition of the excited atomic system formed by the quantised black hole modes and the incoming modes. 

\section{Conclusion}
Laboratory models for black holes and Hawking radiation, serve two purposes. Firstly, they enable us to see in the laboratory, the physics of astrophysical phenomena, which are inaccessible to us experimentally, and secondly,  to test various theoretical ideas associated with quantum gravity. Hawking's original conception of radiation from black holes, does not take into consideration the quantum structure of the horizon of the black hole , advocated by recent work in quantum gravity \cite{rovelli},\cite{beckenstein},\cite{ praznetti}. A complete analysis of such quantum phenomena, requires a full field theoretic study of quantum gravity such as loop quantum gravity \cite{rovelli}. Toy models , such as the one we have studied, give clues to how Hawking radiation would be modified if black hole modes are quantized. In the spirit of  Beckenstein and Mukhanov \cite{beckenstein}, who made the proposal of treating black holes  as quantum atomic systems and considering  the resultant modification in the character of thermal Hawking radiation, we have tried to see how this can be realized in a quantum optical system that mimics such a proposal.  Firstly, we have considered the fact that a two-mode squeezed sytem of a parametric amplifier, can give rise to a "Hawking-like" distribution for the outgoing (signal) modes, this corresponds to the standard Hawking picture, where the black hole is considered to be classical. Then, we have included the quantization of the black hole modes to show that, indeed, the thermal character  of the Hawking radiation, is  modified. Moreover we have shown how the entanglement properties change without and with quantization of the black hole modes. Indeed, the figures for the entropy of the black hole modes (Fig.\ref{fig5:trlentropya})and that for the outgoing modes shown in (Fig.\ref{fig8:trlentropyb}), are suggestive of a final stable state for the black hole entangled with the radiation measured at infinity.

The Hamiltonian that we have considered can be viewed in two different ways, one of which admits a quadratic polynomially deformed su(1,1),  while the other admits a quadratic polynomially deformed su(2) spectrum generating algebras. 
The utilization of quadratic polynomial deformations of su(1,1)  algebras  has enabled us to study the entanglement properties of the three mode quantum system (incoming modes, black hole modes and the outgoing modes) in the long time limit, in an exact fashion.  
We find that the system has superradiant  properties that are akin to atomic "Dicke superradiance" by means of a mapping of the system to a quadratic polynomially deformed su(2) algebra. 

Since, we can control quantum optical systems and atom-radiation systems (cavity systems) are easily created, modification effects on the Hawking radiation due to black hole quantization can be actually realised in this  laboratory analogue. We hope that this study of a simple analogue would help in giving some direction to the complete field theoretic studies in quantum gravity and also help in establishing the similarity in the  behaviour of superradiance from atomic systems and superradiance from black holes. In fact, in any canonical quantum gravity study, ultimately a mode decomposition in terms of creation and annihilation operators of the field is done, so the symmetries, in the form of deformed polynomial algebras, could prove to be useful for more comprehensive studies. Our ideas could also be applied to other type of black hole analogue systems such as those using Bose Einstein Condensates. Such studies, we feel, are worth pursuing, and we intend to follow these in our future work.
\section{Acknowledgements}
Bindu Bambah would like to thank Mr.M Naveen Kumar for assistance in plotting some of the figures. 
T. Shreecharan would like to thank the UGC for a D.S. Kothari postdoctoral fellowship.
K. Siva Prasad would like to thank the UGC-CAS for financial support under its project fellowship scheme.

\end{document}